\documentclass[prl,aps,twocolumn,groupedaddress,floats,showpacs]{revtex4}
\usepackage{graphicx}
\usepackage{dcolumn}
\usepackage{bm}
\usepackage{color}
\usepackage{amsfonts}
\usepackage{mathtools}
\usepackage{amssymb}
\usepackage{hyperref}
\begin{document}

\newcounter{fnnumber}

\newcommand{\bea}{\begin{eqnarray}}
\newcommand{\eea}{\end{eqnarray}}
\newcommand{\be}{\begin{equation}}
\newcommand{\ee}{\end{equation}}
\newcommand{\bes}{\begin{equation*}}
\newcommand{\ees}{\end{equation*}}
\newcommand{\ds}{\displaystyle}
\newcommand{\rr}{\mathbf{r}}
\newcommand{\kk}{\mathbf{k}}
\newcommand{\pp}{\mathbf{p}}
\newcommand{\qq}{\mathbf{q}}
\newcommand{\ra}{\rangle}
\newcommand{\la}{\langle}
\newcommand{\si}{\sigma}
\newcommand{\sip}{{\sigma'}}
\newcommand{\up}{\uparrow}
\newcommand{\down}{\downarrow}
\newcommand{\GtN}{\tilde{G}_{\mathcal{N}}}
\newcommand{\Gt}{\tilde{G}}
\newcommand{\Nr}{\mathcal{N}}
\newcommand{\Dr}{\mathcal{D}}
\newcommand{\Cr}{\mathcal{C}}
\newcommand{\bL}{L} 
\renewcommand{\Pr}{\mathcal{P}}
\newcommand{\Sr}{\mathcal{S}}
\newcommand{\blue}{\color{blue}}

\author{Riccardo Rossi\footnote{\href{mailto:riccardo.rossi@ens.fr}{riccardo.rossi@ens.fr}}}

\affiliation{Laboratoire de Physique Statistique, \'Ecole Normale Sup\'erieure, 
 75005 Paris, France}
 

\title{Determinant Diagrammatic Monte Carlo in the Thermodynamic Limit}

\begin{abstract}
  We present a simple trick that allows to consider the sum of all connected Feynman diagrams at fixed position of interaction vertices for general fermionic models, such that the thermodynamic limit can be taken analytically. With our approach one can achieve superior performance compared to conventional Diagrammatic Monte Carlo, while rendering the algorithmic part dramatically simpler. By considering the sum of all connected diagrams at once, we allow for massive cancellations between different diagrams, greatly reducing the sign problem. In the end, the computational effort increase only exponentially with the order of the expansion, which should be contrasted with the factorial growth of the standard diagrammatic technique. We illustrate the efficiency of the technique for the two-dimensional Fermi-Hubbard model.
\end{abstract}

\pacs{71.10.Fd, 02.70.Ss}

\maketitle
Finding an efficient method to solve the quantum many-body problem is of fundamental physical importance. A promising strategy to gain insight is represented by quantum simulators. For example, experimentalists working with cold atoms in optical lattices are able to realise one of the most prominent models of strongly-correlated electrons, the Hubbard model. A study of its equation of state at low-temperature has been reported in~\cite{Cocchi_16}. Moreover, short-range antiferromagnetic correlations have been observed~\cite{hart_15,Greif_15, Parsons_16, Cheuk_16, Drewes_16}, and, very recently, even a long-range antiferromagnetic state was realized~\cite{antiferro_exp_16}.

    From the theoretical side, making predictions for strongly-correlated fermionic systems is a very challenging problem. Quantum Monte Carlo methods are affected by the infamous sign problem when dealing with fermionic models, which in general precludes reaching low-temperature and large system sizes (see~\cite{Troyer_05, leblanc_15} and references therein). The simulation time to obtain a given precision scales exponentially with the system size and the inverse of the temperature. In the strongly correlated regime, it is then very difficult to extrapolate to the Thermodynamic Limit (TL). However, the fermionic sign gives an unexpected advantage when considering fermionic perturbation theory. For bosonic theories, typically, the perturbative expansion has a zero radius of convergence. This is due to the factorial number of Feynman diagrams all contributing with the same sign. For fermions, strong cancellations between different diagrams at fixed order lead in general to a finite convergence radius on a lattice at finite temperature. This happens because for sufficiently small interactions of whatever sign (or phase) the system is stable on a lattice at non-zero temperature. This was seen numerically~\cite{van_houcke_10, kozik_10}, and even proved mathematically for the Hubbard model at high enough temperature~\cite{Mastropietro}. Perturbation theory is therefore a powerful tool to study fermionic theories. By analytic continuation, any point of the phase diagram which is in the same phase as the perturbative starting point can be reached in principle, provided that one has a method to compute the diagrammatic series to high order. At the moment, there are two numerical methods to evaluate diagrammatic expansions at high order, Determinant Diagrammatic Monte Carlo (DDMC), and Diagrammatic Monte Carlo (DiagMC). Major recent achievements of DiagMC are the study of the normal phase of the unitary fermi gas~\cite{van_houcke_12}, the determination of the ground state phase diagram of the Hubbard model up to filling factor 0.7 and interactions $U/t\le 4$~\cite{Deng_15}, and the settlement of the Bose-metal issue~\cite{Gukelberger_14}. Given the power and the versatility of the technique, systematic diagrammatic extensions of Dynamical Mean Field Theory are now being studied~\cite{Iskaskov_16, Gukelberger_16}.

    In DiagMC~\cite{van_houcke_10, kozik_10, kulagin_13, mishchenko_14} one expresses the perturbative expansion in terms of connected Feynman diagrams, which can be written directly in the TL. One then performs a random walk in the space of topologies and of integration variables of Feynman diagrams. In DDMC~\cite{bourovski_04, rubtsov_05, burovski_06} one computes finite-volume perturbative contributions, considering at once the sum of all Feynman diagrams (connected and disconnected) at fixed position of interaction vertices. One lets the interaction vertices perform a random walk in the space-time simulation volume. In principle, in order to compute the sum of all Feynman diagrams, one should consider all possible connections between these vertices, of which there are a factorial number. However, it has been pointed out that all these fermionic permutations can be grouped in determinants, which can be computed in polynomial time. In addition, summing all diagrams together already accounts for massive cancellations. For example, the Hubbard model at half filling has no sign problem within DDMC, because the sum of all diagrams at a given order is positive definite for every vertex configuration; in contrast the individual diagrams considered by DiagMC have positive and negative signs. The major downside of DDMC is that one has to consider a finite-size system, and this is a serious problem in the generic case where one has sign problem. Intensive quantities, like the density or the energy per site, are computed in DDMC as the ratio of two quantites which increase exponentially with the volume.
    
    In this Letter we present a way to compute directly these intensive quantities, such that the thermodynamic limit can be taken analytically. It is then clear that the sign problem is not present, at least not in the traditional form. As we consider the sum of all connected Feynman diagrams at once, we already account for cancellations between different diagrams at the same order, greatly reducing the variance in the sampling. In the end, the total computational cost increases only exponentially with the order, which is to be compared with the factorial increase of the standard diagrammatic technique. We provide numerical proof of the efficiency of the technique by applying it to the two-dimensional Hubbard model at low temperature and weak coupling, where the series is fast converging. We compute the radius of convergence of the series, which is determined by a phase transition happening for negative values of the interaction.

\begin{figure} 
  \includegraphics[width=0.48\textwidth, height=0.4\textwidth]{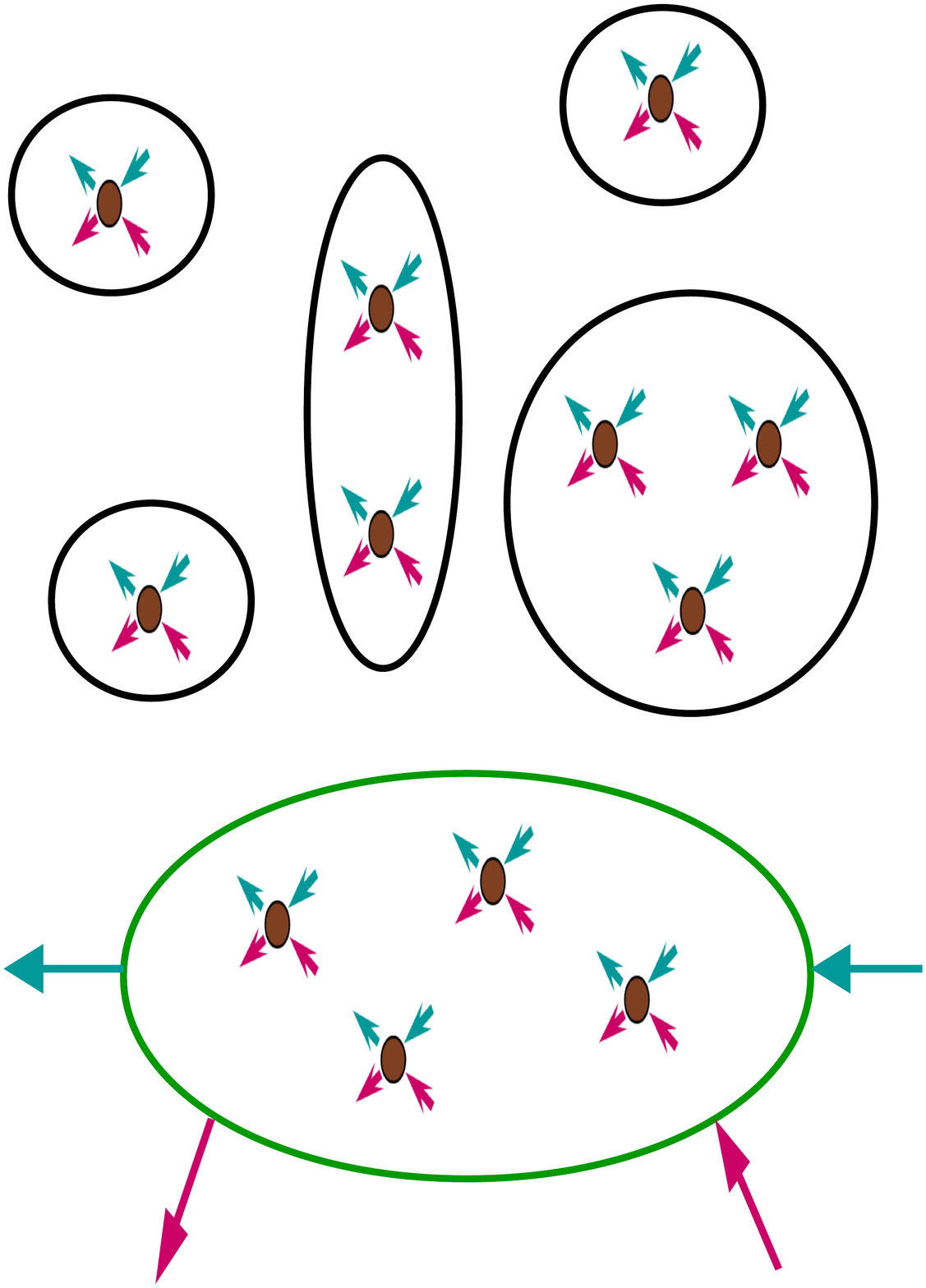}
     \caption{\label{connected_diag} A twelve-order disconnected contribution for the four-point correlation function. The big arrows represent the external lines, while the points with small arrows represent two-particle interaction vertices. The ellipses represent connected parts. The green ellipse represents all the diagrams contributing to the connected part with four interaction vertices at fixed space-time position. The black ellipses can be considered collectively as arising from connected+disconnected diagrams with no external lines.}
\end{figure}

Let us present an intuitive diagrammatic derivation of the analytical result of this work. As pointed out in~\cite{bourovski_04, rubtsov_05, burovski_06}, we can express the sum of all diagrams with fixed space-time position of interaction vertices in terms of determinants. This is due to the fact that a determinant accounts for all the possible connections between vertices, with the right sign for fermions. In this way it is clear that one generates all diagrams, connected and disconnected. We would like to remove disconnected diagrams, as we know that only connected diagrams will contribute to intensive quantities~\cite{abrikosov_book}. Let us consider a disconnected diagram. It can be divided in a part which is connected to the external points of the function we are looking at, and another part which is not connected to it (see Figure \ref{connected_diag}). This correspondence is one-to-one. It is then clear that we can write a recursive formula for the connected part: we subtract from the set of all diagrams (connected or not) those which can be divided in a connected part and a disconnected part. Let $c_E(V)$ be the sum of connected diagrams contributing to a correlation function with a set of external points represented by $E$, and with interaction vertices $V=\{v_1,\dots,v_n\}$ at fixed space-time position. Similarly, let $a_E(V)$ be the sum of all diagrams, connected and disconnected. In particular, $a_\emptyset(V)$ denotes all diagrams with interaction vertices $V$ and no external points (they are the diagrams contributing to the grand-partition function). $a_{E'}(V')$ is easy to compute for every $E'$ and $V'$ as it can be expressed in terms of determinants, but we are interested in obtaining $c_E(V)$. If we know $c_E(S)$ for all $S$ proper subset of $V$, we can compute $c_E(V)$ from
\begin{equation}\label{recursive_formula}
c_E(V)=a_E(V)-\sum_{S\subsetneq V}c_E(S)\;a_{\emptyset}(V\setminus S)\; .
\end{equation}
If one wants to compute the grand-canonical free energy (the pressure for an homogeneous system), which has no external points, one has to render one of the vertices ``special'' and consider connectedness with respect to it. If the vertices are indistinguishable, in order to obtain the $n$-th order diagrammatic contribution for the correlation function one has to divide by $n!$ after summation of $c_E(V)$ over space-time position of $v_1,\dots,v_n$. The integral of $c_E(V)$ over the space-time positions of the $v_1, \dots, v_n$ is convergent, in other words, one can consider directly the TL where the positions of the interaction vertices is unconstrained. The integration over the space-time position of the vertices is performed with standard Markov chain Monte Carlo algorithm. We sample a linear combination of the order $n$ and the order $n-1$ for normalization purposes, but alternatively one could use order-changing updates as in usual DDMC and DiagMC implementations.

  \begin{figure}\hspace*{-17pt}
  \includegraphics[width=0.58\textwidth]{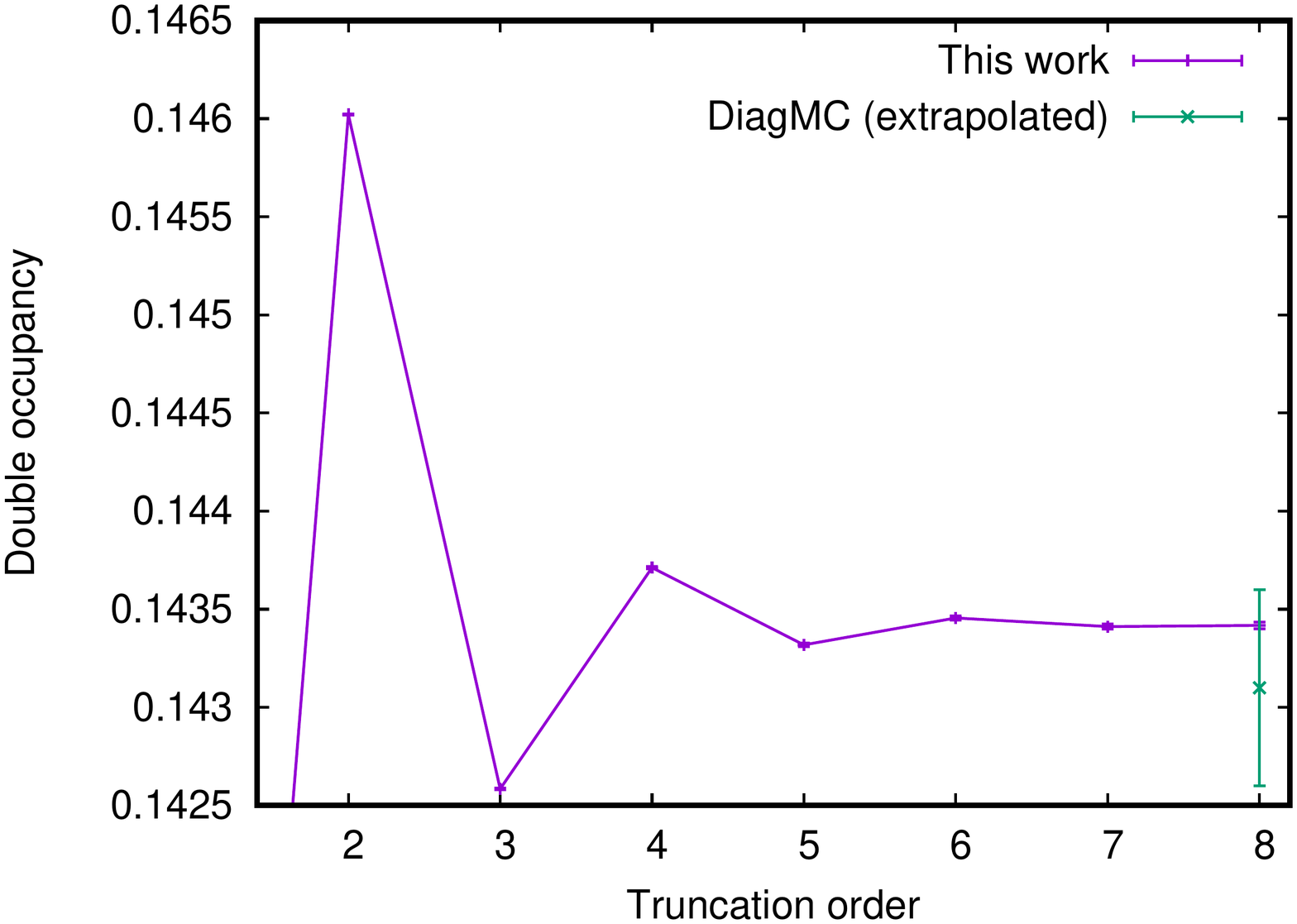}
     \caption{\label{D_tldmc} Double occupancy as a function of truncation order, for $\beta=8$, $U=2$ and $n=0.87500(2)$. The DiagMC number is taken from~\cite{leblanc_15} and it refers to infinite-order extrapolation of the bold series for $n=0.875$.}
\end{figure}

  Let us discuss how the computational cost varies with the order $n=|V|$. The time to compute $a_E(S)$ and $a_\emptyset(S)$ for all $S\subset V$ scales like $n^3\,2^n$ ($n^3$ is the cost to compute the determinant of a $n\times n$ matrix, and we have to compute $2^n$ determinants roughly of this size). It can  be shown that the number of arithmetic operations needed to get $c_E(V)$ from the recursive formula~\eqref{recursive_formula} is proportional to $3^n$, which is the main contribution for $n$ large enough. The observables we sample are not sign definite, this is reminiscent of the fermionic sign problem. The crucial advantage of this technique over DiagMC is that we eliminate topologies from configuration space. In the case where all diagrams have the same sign, this is not a big advantage. However, in fermionic models this is not at all the case, there is almost perfect cancellation between different diagram topologies occuring with opposite signs. The cancellations are so strong that the sum over the diagrams at a certain order divided by the same sum taken with absolute values goes to zero factorially like the number of diagrams, this can be seen as a consequence of having a finite radius of convergence. This means that if one samples topologies one-by-one, like in DiagMC, a factorial ``sign problem'' is encountered (see~\cite{sign_problem_solved} for a more detailed discussion). We see therefore that the trick of summing over all connected topologies allows to greatly alleviate this reminiscence of the sign problem, leaving us with a ``sign problem'' from the integration over the space-time positions of interaction vertices that increases at most exponentially with the number of vertices.
  \begin{figure}\hspace*{-17pt}
  \includegraphics[width=0.58\textwidth]{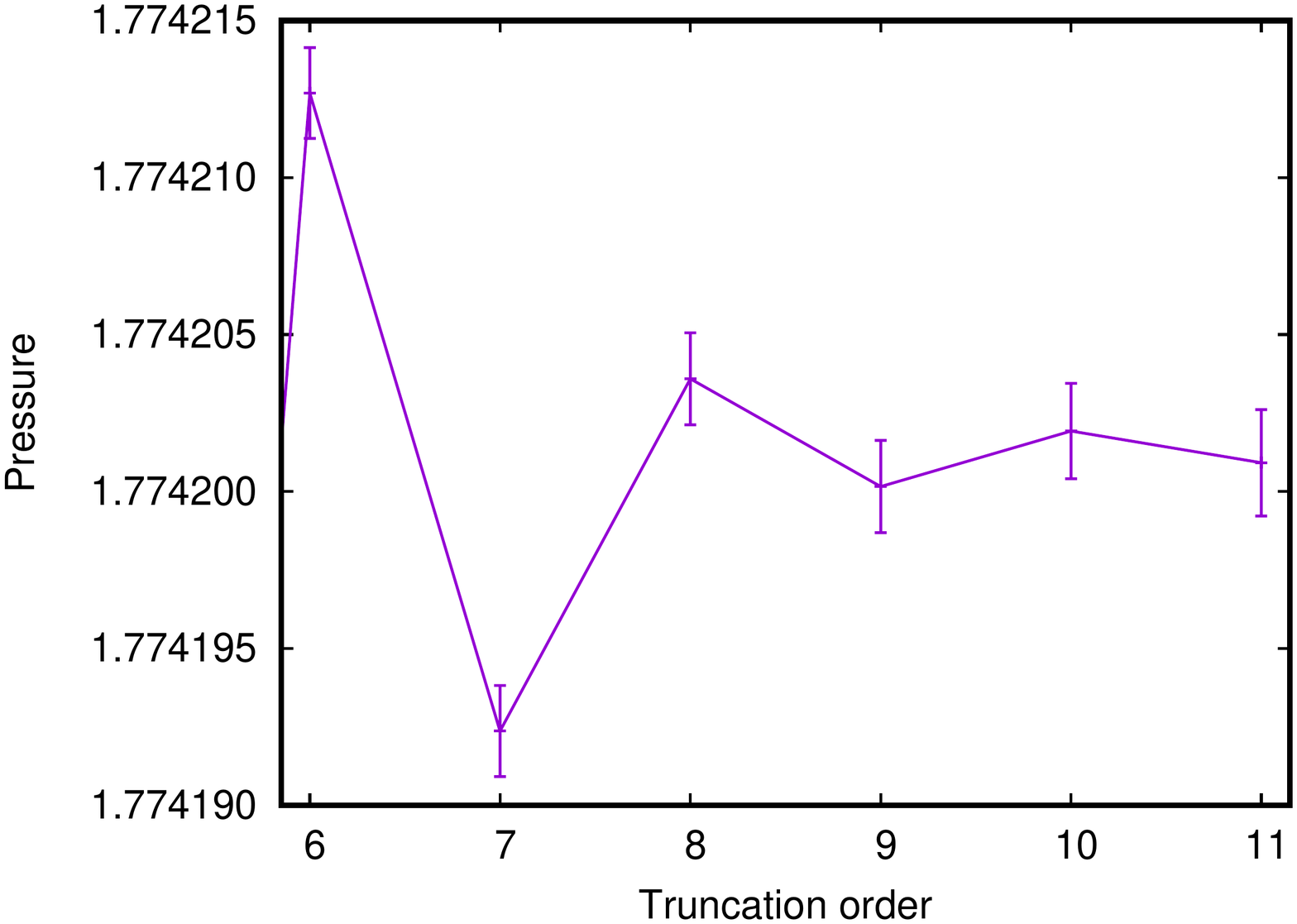}
  \caption{\label{P_tldmc} Pressure as a function of truncation order, for $\beta=8$, $U=2$, and $n=0.87500(2)$.
    }
\end{figure}
  One might wonder if paying an exponential cost to remove disconnected topologies is really worth it, as we could compute the sum of {\it all} topologies in polynomial time, as it is done in DDMC. The advantage of considering only connected diagrams is that we do not suffer from the traditional form of sign problem, that is, the prohibitive scaling of computational time with system size. An analogous situation was found in the context of out-of-equilibrium impurity models~\cite{Parcollet_15}, where in order to consider the long time evolution it was found advantageous to pay an exponential cost for each Monte Carlo step to explicitely eliminate disconnected diagrams\footnote{In this context one starts from a non-interacting state and let the system evolve in real time. Observables are computed in power series of the interaction potential, whose coefficients can be expressed in terms of determinants. The use of the recursive formula for eliminating disconnected diagrams is not necessary here, because the sum over the Keldish indices directly eliminates disconnected diagrams.}. For these reasons, we are able to reach higher orders than DiagMC, even without resumming classes of diagrams more complicated than tadpoles (for the Hubbard model DiagMC arrives at order $\sim$ 6 for both bare and bold series). Unlike DDMC, sign problem does not limit us to work at half filling or with attractive interactions.

\begin{figure}\hspace*{-17pt}
  \includegraphics[width=0.58\textwidth]{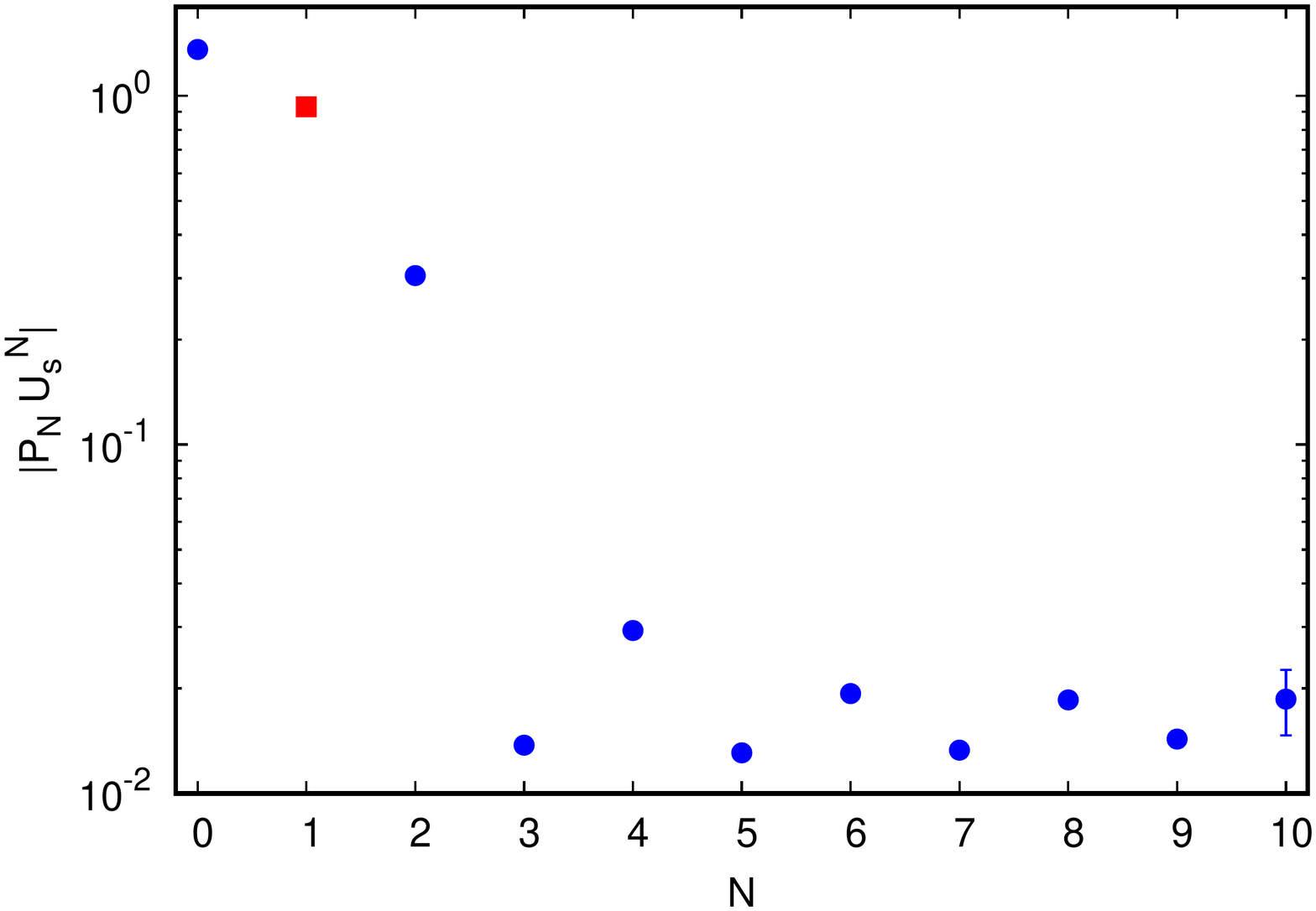} 
  \caption{\label{P_radius} $|P_N\;U_s^N|$ as a function of the expansion order $N$, where $U_s=-5.1$ and $P_N$ is the coefficient of the series for the pressure, $P(U)=\sum_{N=0}^\infty P_N\;U^N$. The blue circle points correspond to positive values of $P_N U_S^N$, while the red square point correspond to a negative value. We see that starting from order $\sim 5$ we are in the high-order asymptotic regime. This implies that the radius of convergence of the series is $R=5.1(1)$.
    }
\end{figure}

  We now discuss the results obtained by implementing this method for the two-dimensional Hubbard model. Without loss of generality we can set the hopping parameter $t$ to one. We consider inverse temperature $\beta=8$, repulsive on-site interaction parameter $U=2$, at density $n=0.87500(2)$ near to half-filling. All our error bars correspond to one standard deviation. We resum all bare tadpoles diagrams, whose effect is to shift the chemical potential $\mu(U)=\mu_0+Un_0/2$, where $\mu_0$ is the chemical potential needed to get the density $n_0$ in the absence of interactions (this corresponds to the first-order semi-bold scheme introduced in~\cite{rossi_15_04}). This is useful because one has a smaller density shift as a function of $U$. We compare thermodynamical quantities with DiagMC benchmarks from~\cite{leblanc_15}. We find compatible results, with error bars one order of magnitude smaller (see Figure~\ref{D_tldmc}). We estimate the chemical potential at fixed density $n=0.875$ to be $\mu=0.55978(7)$, while DiagMC gives $\mu_{Diag}=0.558(3)$. For the energy per site, we have $E=-1.25992(6)$, while $E_{Diag}=-1.2600(6)$. As the entropy is relevant for experiments in optical lattices, we also give the value of the entropy per site $S=0.1958(4)$. We have pushed the computations up to eleven orders for the pressure (let us note that for the pressure we have one order more for free) as the error bars continued to stay bounded (see Figure~\ref{P_tldmc}), spending seven thousand CPU hours. Let us only remark here that the exponential cost to go to higher orders is compensated by an exponential convergence as a function of the order for a convergent series, resulting in an error bar that decays as a power law as a function of computer time~\cite{sign_problem_solved}. We have estimated the radius of convergence of the series in $U$ by looking at coefficients, see Figure~\ref{P_radius}. This is compatible with a phase transition happening at $U=-5.1(1)$, when the system is electron-doped $n>1$. This is in qualitative agreement with the established phase diagram for the attractive Hubbard model~\cite{Paiva_04}. The subleading alternating structure shown in Figure~\ref{P_radius} is compatible with an additional singularity around $U=6$, but further work is needed to rule out other possibilities. In principle, the series converges exponentially for $|U|<5.1$. In practice, in order to get accurate results for moderate values of interactions it is necessary to perform analytical continuation/resummation of the series in order to accelerate (or extend) the convergence.
  
  We stress that there is no fundamental difference in the sampling between different physical quantities, as we expect the external points to play a minor role at high-enough orders. In the high-order limit, external points can be thought as ``boundary terms'', the bulk being the large number of internal interaction vertices. This intuition is supported by the universality of large-order behaviour of perturbation theory, which is independent of the configuration of the external points~\cite{zinn_justin_book}. Nevertheless, there exists tricks that can improve numerical sampling in a decisive way. Let us briefly discuss a couple of them for the Green's function. The simplest modification is to consider the unlegged Green's function, which can be further divided in a fully-dressed tadpole diagram, and another part. This has the advantage to make possible to obtain statistics for all values of the Green's function at once, as the diagrams depend on the external points only through these legs in the space-time representation, obtaining as well two orders more for free. To go further, one has to sample the self-energy. We want to get rid of one-particle reducible diagrams. A general one-particle reducible diagram can be divided in a part which is one-particle irreducible with respect to one of the external points, and another part consisting of a diagram for the Green's function. We can then derive a recursive formula for the self-energy of the same form as equation~\eqref{recursive_formula}.

Finally, we note that it is possible to generalize the method to more general diagrammatic schemes by the use of a shifted action~\cite{rossi_15_04}, where one has to consider additional interaction vertices that act as counterterms.

In conclusion, we have presented an elementary yet efficient method for the computation of high-order perturbative expansions of general fermionic models directly in thermodynamic limit. Formally, this method has superior algorithmic properties than DiagMC. We have verified the formal arguments by providing numerical proof for the low-temperature Hubbard model at small interaction, obtaining results that are the state of the art in this regime. We have computed the radius of convergence of the series, which shows that the bare series is convergent up to moderate value of interaction potential. Moreover, we have shown that achieving such high orders can be used to detect singularities of thermodynamical functions, which are known to indicate phase transitions.

A natural continuation of this work would be the study of the pseudo-gap regime of the two-dimensional Hubbard model, where the new algorithm has the potential to extend to lower temperature the DiagMC results of~\cite{diagMC_pseudo}.\\
    
I am grateful to Kris Van Houcke and F\'elix Werner for useful suggestions, constant support of the project, and help with writing the manuscript.
I thank Boris Svistunov for fruitful discussions and for proposing improvements of the manuscript. I acknowledge useful discussions with Tommaso Comparin. This work has  been  partially supported  by  the  Simons  Foundation  within the  Many  Electron  Collaboration  framework. Simulations ran on the cluster {\it MesoPSL} (Region Ile-de-France / grant
ANR-10-EQPX-29-01). I am also affiliated to {\it CNRS, Universit\'e de recherche Paris Sciences et Lettres}, {\it UPMC}, {\it Universit\'e Paris Diderot}, {\it Sorbonne Universit\'es}, and {\it Sorbonne Paris-Cit\'e}.
b\bibliography{biblio} 
\end{document}